\documentclass[9pt,twocolumn,twoside]{optica}
\setboolean{shortarticle}{false}
\setboolean{minireview}{false}

\usepackage{upgreek}
\hypersetup{
    colorlinks=true,
    linkcolor=red,
    citecolor=blue,
    filecolor=magenta,
    urlcolor=magenta,
}

\usepackage{siunitx}
\DeclareSIUnit{\belmilliwatt}{Bm}
\DeclareSIUnit{\dBm}{\deci\belmilliwatt}

\usepackage{multirow}
\usepackage{tabularx}
\usepackage{ulem}
\usepackage{fancyhdr}
\usepackage{array}
\usepackage{booktabs}
\usepackage{color}
\usepackage{bm}
\usepackage{eurosym}
\usepackage{subfigure}
\usepackage{comment}
\usepackage{scrhack}
\usepackage{physics}
\usepackage{balance}
\usepackage{float}
\usepackage{flushend}
\title{Bidirectional electro-optic conversion reaching 1\% efficiency with thin-film lithium niobate}

\author[1,$\dagger$]{Yuntao Xu}
\author[1,$\dagger$]{Ayed Al Sayem}
\author[1,$\ddagger$]{Linran Fan}
\author[1]{Sihao Wang}
\author[1]{Risheng Cheng}
\author[1]{Chang-Ling Zou}
\author[1]{Wei Fu}
\author[1]{Likai Yang}
\author[1]{Mingrui Xu}
\author[1,*]{Hong X. Tang}

\affil[1]{Department of Electrical Engineering, Yale University, New Haven, CT 06511, USA}
\affil[*]{Corresponding author: hong.tang@yale.edu}

\affil[$\dagger$] {These authors contributed equally to this Letter.}
\affil[$\ddagger$]{Current address:  College of Optical Sciences, The University of Arizona, Tucson, Arizona 85721, USA}

\begin{abstract}
Superconducting cavity electro-optics (EO) presents a promising route to coherently convert microwave and optical photons and distribute quantum entanglement between superconducting circuits over long-distance through an optical network. High EO conversion efficiency demands transduction materials with strong Pockels effect and excellent optical transparency. Thin-film Lithium Niobate (TFLN) offers these desired characteristics however so far has only delivered unidirectional conversion with efficiencies on the order of 10$^{-5}$, largely impacted by its prominent photorefractive (PR) effect at cryogenic temperatures. Here we show that, by mitigating the PR effect and associated charge-screening effect, the device's conversion efficiency can be enhanced by orders of magnitude while maintaining stable cryogenic operation, thus allowing a demonstration of conversion bidirectionality and accurate quantification of on-chip efficiency. With the optimized monolithic integrated superconducting EO device based on TFLN-on-sapphire substrate, an on-chip conversion efficiency of 1.02\% (internal efficiency,  $15.2\%$)  is realized. Our demonstration indicates that with further device improvement, it is feasible for TFLN to approach unitary internal conversion efficiency. 

\end{abstract}

\setboolean{displaycopyright}{true}

\begin{document}

\maketitle

\section{Introduction}

With superconducting qubits recently emerging as one of the most promising platforms for quantum computation \cite{clarke2008superconducting,devoret2013superconducting,kurpiers2018deterministic,chou2018deterministic,zhong2019violating,arute2019quantum} and optical photons being the most suitable quantum information carrier \cite{lvovsky2009optical,o2009photonic}, efficient bidirectional conversion between microwave and optical photons at the quantum level is in critical demand for interfacing distributed quantum computation and quantum communication systems \cite{tsang2010cavity,tsang2011cavity,javerzac2016chip,soltani2017efficient,fan2018superconducting,rueda2016efficient,fu2020ground,eo7_groundstate_fink_2020,mckenna2020cryogenic, holzgrafe2020cavity, youssefi2020cryogenic,blum2015interfacing,gard2017microwave,han2018coherent,petrosyan2019microwave,barzanjeh2012reversible,bochmann2013nanomechanical,andrews2014bidirectional,balram2016coherent,higginbotham2018harnessing,shao2019microwave,forsch2019microwave,han2020cavity,mirhosseini2020quantum,jiang2020efficient,o2014interfacing,williamson2014magneto,hisatomi2016bidirectional,zhu2020waveguide}. Through this microwave-to-optical interconnection, a hybrid system where quantum information is processed by superconducting circuits and distributed with photonic circuits is one of the most promising schemes to build large-scale quantum networks \cite{cirac1997quantum,kimble2008quantum,schoelkopf2008wiring,monroe2014large,zhong2020proposal}. A variety of hybrid platforms have been demonstrated to realize efficient microwave-to-optical conversion, including spin ensembles \cite{blum2015interfacing,gard2017microwave,han2018coherent,petrosyan2019microwave}, electro-optomechanics (EOM) \cite{barzanjeh2012reversible,bochmann2013nanomechanical,andrews2014bidirectional,balram2016coherent,higginbotham2018harnessing,shao2019microwave,forsch2019microwave,han2020cavity,mirhosseini2020quantum,jiang2020efficient}, rare-earth-doped crystal \cite{o2014interfacing,williamson2014magneto}, ferromagnetic magnons \cite{hisatomi2016bidirectional, zhu2020waveguide}, and etc. 
The highest conversion efficiency has been demonstrated in EOM systems using 3D optical cavity and megahertz mechanical membranes \cite{higginbotham2018harnessing}. 

Without relying on intermediate excitations to assist the conversion process, the cavity electro-optics (EO) system directly utilizes the Pockels nonlinearity of the material to realize microwave-to-optical conversion \cite{tsang2010cavity,tsang2011cavity}. In the cavity EO systems, a high frequency microwave resonator is directly coupled to optical resonators without introducing suspended structures or external magnetic fields. With favorable tunability and high optical power handling capability, cavity EO system enables a robust high-efficiency microwave-to-optical frequency conversion as experimentally demonstrated in \cite{fan2018superconducting} at 2K, and recently at the microwave ground state in a dilution refrigerator \cite{eo7_groundstate_fink_2020,fu2020ground}. The highest EO conversion efficiency has been achieved in Aluminium Nitride (AlN) material platform reaching $2\%$ on-chip conversion efficiency. 


The efficiency of cavity EO conversion critically relies on the strength of Pockels nonlinearity and the optical transparency of the transduction materials. Among the few Pockels materials other than AlN, thin-film Lithium Niobate (TFLN) is the most promising candidate because of its excellent optical properties \cite{zhang2017monolithic} along with strong Pockels nonlinearity, which is around thirty times stronger than AlN. Very recently, there have been two successful demonstrations of cavity EO conversion based on TFLN platform \cite{mckenna2020cryogenic,holzgrafe2020cavity}, both realizing impressive vacuum coupling rate ($g_{eo}$) compared to AlN material platform. Nevertheless, the achieved conversion efficiency is on the order of $10^{-5}$. This is largely impeded by the prominent photorefractive (PR) effect which has been particularly challenging for cryogenic LN devices and constrains the sustainable pump power at cryogenic temperatures. Also because of the resonance shift induced by the PR effect at parametric pump powers, only unidirectional microwave-to-optical conversion has been achieved. 

In this article, we demonstrate a hybrid cavity EO converter based on TFLN optical cavities and superconducting niobium nitride (NbN) resonator reaching $1.02 \%$ on-chip efficiency and $15.2 \%$ internal efficiency. The PR effect and charge-screening effects in TFLN are mitigated by a significant margin by incorporating new microwave and optical resonator designs. High conversion efficiency is achieved at elevated pump power levels while maintaining stable device operation at cryogenic temperatures. The improved efficiency and stability not only lead to a demonstration of the desired bidirectional conversion process but also enable the measurement of the converter's full scattering matrix for accurate calibration of the system conversion efficiency. Currently, our device is limited by the use of high-throughput, albeit low-efficient grating couplers and the parasitic loss of microwave resonator due to radio frequency (RF) packaging. With further device improvement in light coupling and cryogenic packaging, we project that the TFLN EO converter could approach unitary internal conversion efficiency and provide a highly competitive transduction platform for future quantum network applications.

\section{TFLN EO converter Design}

\begin{figure}[!t]
\centering
\includegraphics[width=0.48\textwidth]{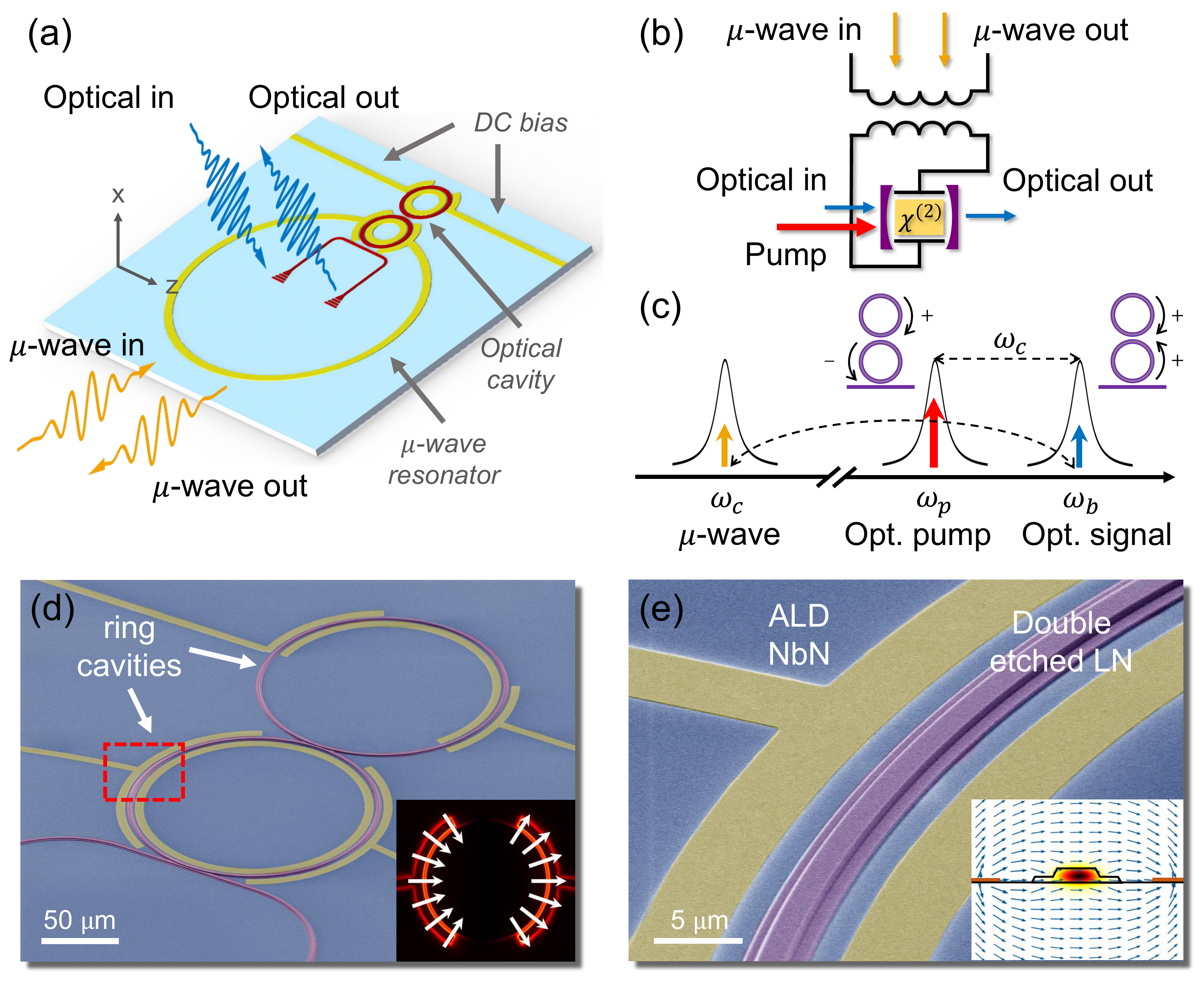}
\caption{\label{fig:schematic}
(a) Schematic layout of the TFLN cavity EO converter, where two strongly coupled microring resonators (red) are co-integrated with an ``Ouroboros'' microwave resonator (yellow). DC electrodes are utilized for on-chip tuning of the optical modes. 
(b) Circuit representation of the cavity EO system. The capacitor plate of the lumped-element microwave resonator is coupled to the $\chi^{(2)}$ optical cavity. 
(c) Three wave-mixing EO conversion process. The optical doublet corresponds to the symmetric and anti-symmetric supermodes of the coupled microrings. A strong pump (red) supported in the anti-symmetric mode parametrically stimulates the bidirectional conversion of signal photons in the symmetric mode (blue) and the microwave mode (yellow) through difference/sum-frequency generation.
(d) and (e) False-color SEM images of the EO converter device. Insets show the electric field profiles of interacting optical and microwave modes.
}
\end{figure}

Figure\,\ref{fig:schematic}(a) illustrates the schematic layout of our TFLN-superconductor hybrid cavity EO converter, which consists of a pair of strongly coupled ring resonators patterned from x-cut TFLN and a superconducting microwave resonator of Niobium Nitride (NbN). The microwave resonator (``Ouroboros''\cite{han2016multimode}) capacitively couples to one of the double rings for EO conversion, while a pair of DC electrodes is coupled to the other microring for electrical tuning of the optical resonance modes. The circuit representation of the cavity EO system is illustrated in Fig.\,\ref{fig:schematic}(b), where the electric field of a lumped-circuit LC resonator overlaps with the optical cavity field within an EO ($\chi^{(2)}$ non-linearity) media through Pockels effect. As a result of the cavity enhanced EO interaction at the triply-resonant condition, the input microwave field modulates the optical pump and produces an optical sideband in the signal mode. Reversely, a microwave field output can be generated by the optical frequency mixing in the LN cavity, and thus realizes the photon-photon quantum frequency conversion~\cite{soltani2017efficient, fan2018superconducting}.

This triply-resonant enhanced EO conversion scheme, as illustrated in Fig.\,\ref{fig:schematic}(c) in frequency domain \cite{fan2018superconducting}, is described by an interaction Hamiltonian
\begin{equation}
\label{eq1}
  H_\mathrm{I}=\hbar g_{eo}(ab^{\dagger}c+a^{\dagger}bc^{\dagger}). 
\end{equation}
Here, $a$, $b$ and $c$ denote annihilation operators for the optical pump, signal and microwave modes, respectively. $g_{eo}$ denotes the vacuum EO interaction strength, which in turn is determined by the mode volume, triple-mode overlap and the Pockels coefficient. The optical frequency doublet (Fig.\,\ref{fig:schematic}(c)) corresponds to the symmetric and anti-symmetric supermodes of the photonic molecule \cite{zhang2019electronically} induced by the strong mutual coupling between the fundamental transverse electric (TE) modes of the double-microring resonators \cite{soltani2017efficient,eo7_groundstate_fink_2020, mckenna2020cryogenic,holzgrafe2020cavity,fu2020ground}. To leverage the highest EO coefficient $r_{33} \sim 30\,\mathrm{pm/V}$ \cite{weis1985lithium} in x-cut TFLN, we utilize the microwave mode with in-plane electrical field. The corresponding microwave and optical mode profiles are shown in the insets of Figs.\,\ref{fig:schematic}(d) and (e). 

To initiate quantum frequency conversion, the lower-frequency supermode in the photonic molecule, i.e. the anti-symmetric optical mode $a$, is loaded with a strong pump to produce a photon-number enhanced electro-optical linear conversion. The advantage of the cavity enhanced coherent conversion is quantified by the cooperativity $C=4 n_p g_{eo}^2 / \kappa_b \kappa_c $, where $n_p$ is the intracavity pump photon number and $\kappa_{b}$, $\kappa_{c}$ are the total energy loss rates for modes $b$ and $c$, receptively. By fine-tuning of the DC bias, the frequency splitting of the optical doublet can be tuned to match the microwave resonance frequency, and fulfill the triply-resonant condition ($\omega_a+\omega_c=\omega_b$). Under this ideal condition, the on-chip peak conversion efficiency is given by, 
\begin{equation}
\label{eq3}
  \eta=\frac{\kappa_{b,ex}}{\kappa_{b}}\frac{\kappa_{c,ex}}{\kappa_{c}}\frac{4C}{(1+C)^{2}}.
\end{equation}
Here, $\kappa_{b,ex}$ ($\kappa_{c,ex}$) is the external loss rates for mode $b$ ($c$), and $\eta_{\mathrm{int}}=4 C/(1+C)^2$ is the internal conversion efficiency. Optimal conversion is achieved when $C=1$, and the corresponding on-chip efficiency is eventually limited by the extraction ratio $\kappa_{b,ex}\kappa_{c,ex}/\kappa_{b}\kappa_{c}$.

\section{PR effect mitigation at cryogenic temperature}

\begin{figure}
\centering
\includegraphics[width=0.48\textwidth]{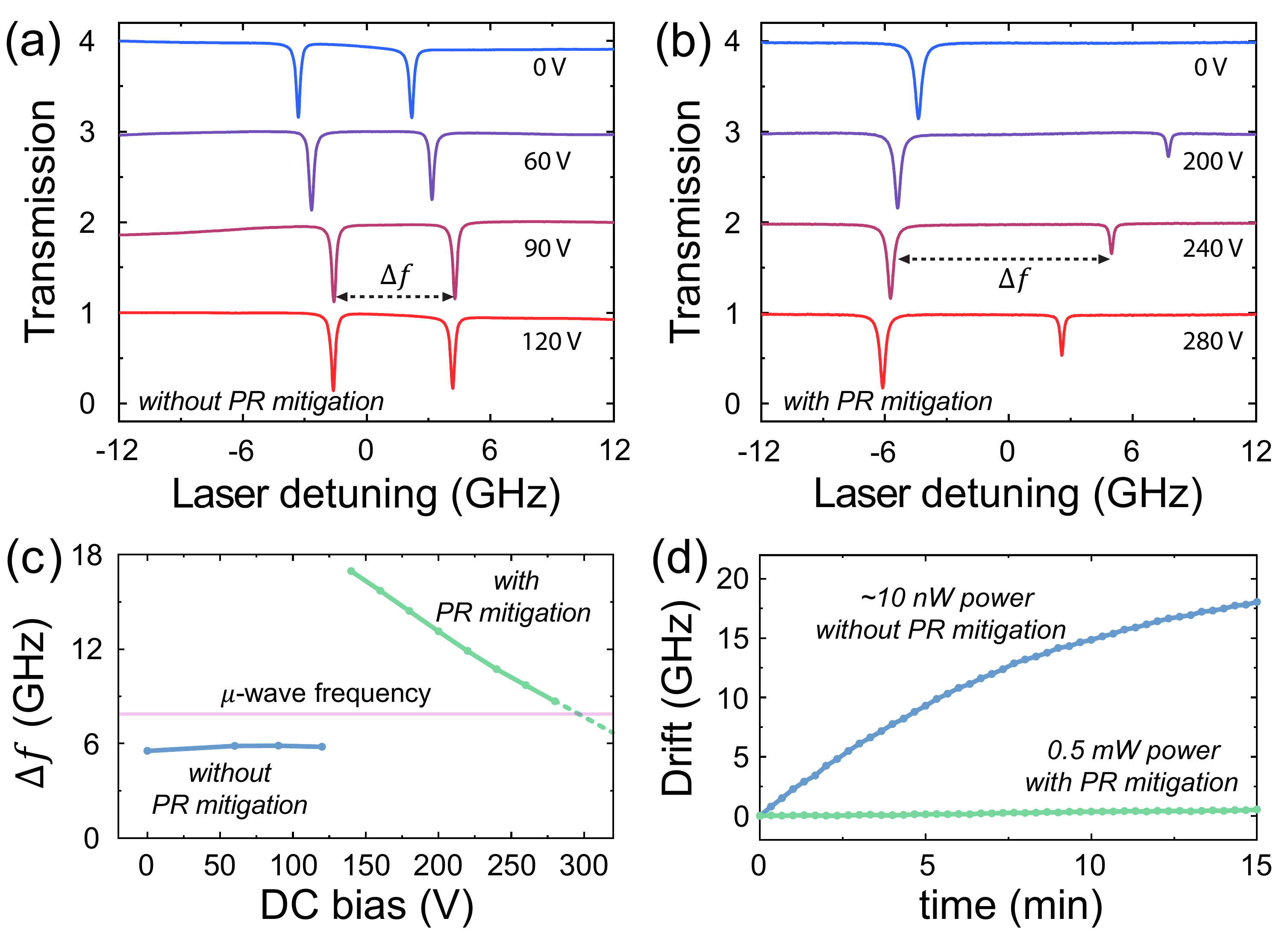}
\caption{\label{fig:PR} (a) Laser-scan spectra of the optical resonances under charge-screening effect. After illumination the supermode frequency splitting $\Delta f$ remains unchanged despite of varying DC bias. (b) Laser-scan spectra of optical resonances after PR effect mitigation. The DC tuning effect is maintained after illumination, with no apparent charge-screening being observed. (c) The supermode splitting $\Delta f$ as a function of DC bias with and without PR mitigation. (d) The resonance frequency shift under light illumination with and without PR mitigation. 
}
\end{figure}

As indicated by Eq.\,\ref{eq3}, the EO conversion efficiency $\eta$ is proportional to pump photon number. One of the major obstacles hindering high conversion efficiency on TFLN platform is the strong PR effect and associated charge-screening effect, which limit achievable intracavity pump power~\cite{mckenna2020cryogenic,holzgrafe2020cavity}. The PR effect in LN is a combination of cascaded processes that first builds up a space-charge field in presence of light illumination and subsequently modulates the refractive index of the material~\cite{weis1985lithium}. With the PR effect, highly dynamic resonance shift prevents strong pump light from launching into a stable resonance. Especially at cryogenic temperature, the relaxation time of the PR effect is found to be extremely long, leading to an accumulated space-charge field that modulate the resonance frequency~\cite{xu2020photorefractioninduced}. Additionally, the space-charge field induced by the PR effect can screen and even cancel the external DC bias voltage, therefore disturb the alignment of resonances~\cite{holzgrafe2020toward}. An illustration of this charge-screening effect is shown in Fig.\,\ref{fig:PR}, where DC bias is applied to a LN double-microring resonator to tune the supermode splitting. After the laser scans across the optical resonances for several times, the DC tuning becomes ineffective and the supermode splitting $\Delta f$ restores to its zero-bias value. The PR effect and its associated charge-screening effect not only limit the intracavity pump power, but also hinder future quantum applications that require long-time stable operation.

We found the PR effect and the charge-screening effect are highly related to the fabrication process. While the underlying mechanism remains to be investigated, we identified that the commonly used intermediate layer of silicon dioxide ($\mathrm{SiO_2}$) layer deposited on TFLN aggravates these photo-induced charge effects. The details of these findings and the room temperature PR effect characterizations can be found in Ref.~\cite{xu2020mitigating}. The effectiveness of this PR effect mitigation approach is also manifested at cryogenic temperature, where the optical cavities without $\mathrm{SiO_2}$ cladding experience significantly reduced PR and charge-screening effect. With this mitigation approach, the DC tuning can be maintained during illumination (Fig.\,\ref{fig:PR}(b)), making it feasible to fine tune the optical frequency difference to match the microwave resonance frequency (Fig.\,\ref{fig:PR}(c)). As the frequency difference between two supermodes is stable during operation, no further feedback on DC bias is required, which greatly simplifies the measurement. The absolute resonance shift induce by the PR effect is also reduced by a significant margin, as shown in Fig.\,\ref{fig:PR}(d). As a result, triple-resonance condition can be easily obtained for boosted intracavity intensities after compensating the weak PR-induced drift by a slow PID feedback on drive laser wavelength.

\section{Device fabrication and cryogenic characterization} 

Here we briefly outline the device fabrication process, which starts with \SI{600}{\nano\meter}-thick TFLN bonded on a sapphire substrate (supplied by NANOLN). Figures\,\ref{fig:schematic}(d) and (e) show the false-color scanning electron microscope (SEM) images of the fabricated EO converter. All structures are patterned with electron beam lithography (EBL) using Hydrogen silsesquioxane (HSQ) resist. Low-loss ridge waveguides and microrings are defined by etching \SI{350}{\nano\meter} into TFLN by $\mathrm{Ar^+}$ milling. The microring resonators have a radius of \SI{80}{\micro\meter} and a top width of \SI{1.6}{\micro\meter}. As shown in Fig.\,\ref{fig:schematic}(e), the majority of remaining \SI{250}{\nano\meter} TFLN slab is removed in a second $\mathrm{Ar^+}$ milling step in order to minimize the total coverage of TFLN, leaving only a \SI{3.5}{\micro\meter} wide pedestal to support the microring resonators. This is a crucial step to suppress the microwave loss induced by the strong piezo-electricity of LN. After patterning the optical structures, superconducting NbN is deposited on the sapphire substrate using atomic layer deposition (ALD), followed by a fluorine-based etching to define the microwave resonator and tuning electrodes. The microwave resonator layout is adjusted to avoid directly crossing over the uncladded optical structures.

\begin{figure}[t]
\centering
\includegraphics[width=0.48\textwidth]{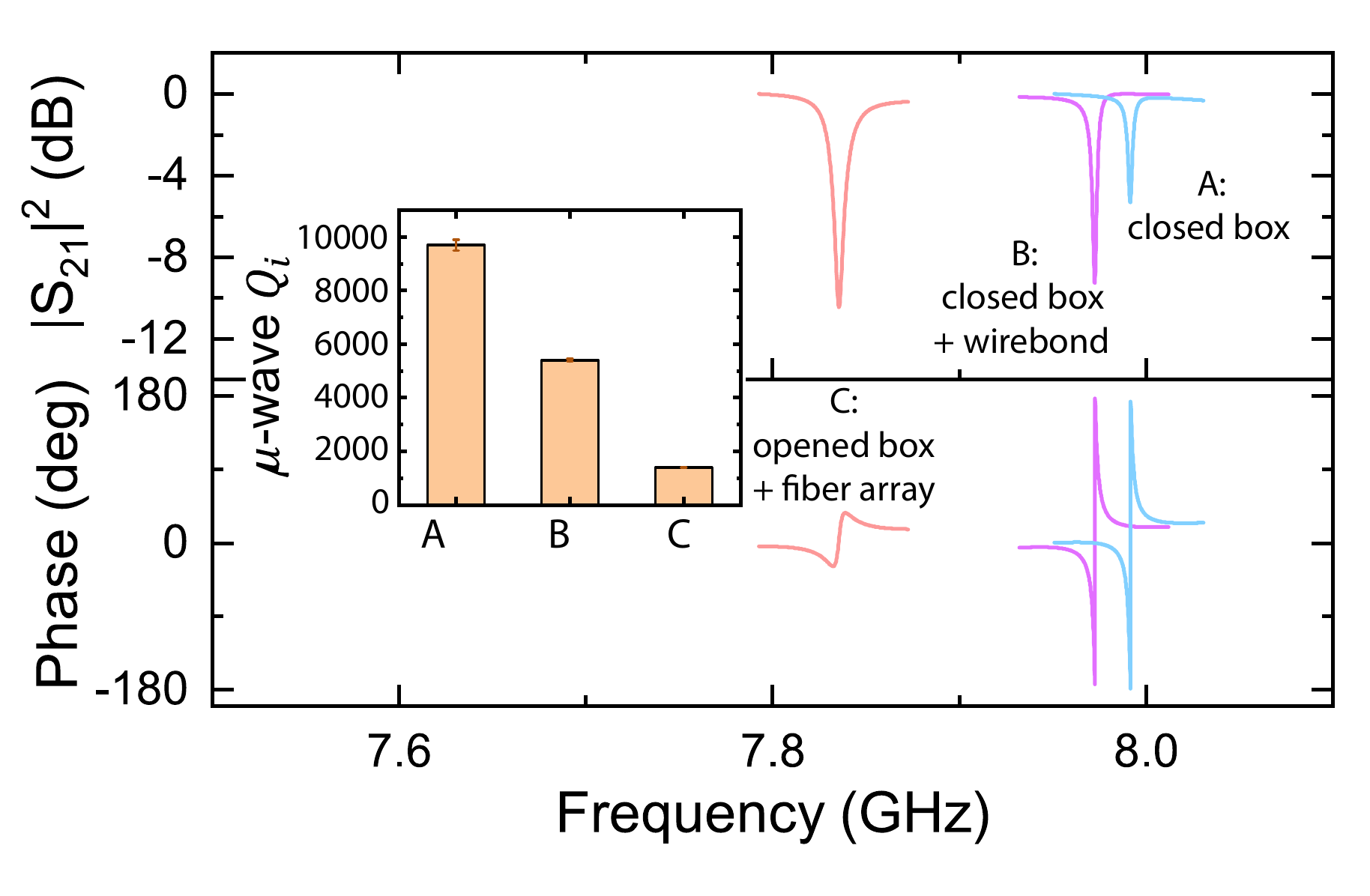}
\caption{\label{fig:MMQ} Microwave reflection spectrum of the device (A) in a closed copper box (B) in a closed copper box with wire bonds (C) packaged with fiber-array for conversion measurement. The inset shows the fitted intrinsic microwave quality factors.} 
\end{figure}

Prior to cryogenic packaging, the as-fabricated device dies are characterized separately in microwave and optical domains. A pair of grating couplers designed for TE polarized light is used to couple light in and out the device. The average intrinsic quality (Q) factor of the fundamental TE modes of the microring resonator is measured to be $\sim 9\times10^5$.
For the microwave readout, we utilize a coaxial cable terminated with a hoop antenna~\cite{han2018coherent, fu2020ground}. While the optical transmission can be measured on die-level at room and cryogenic temperatures showing no apparent change in performances, die-level microwave Q testing is more involved because it has to be accomplished at cryogenic temperatures. The superconducting resonator is first characterized in a RF tight, compact cooper box package to suppress the undesired microwave modes from coupling to the on-chip resonator. Under this condition, the device exhibits an intrinsic Q of $9.7 \times 10^3$ at \SI{2.6}{\kelvin} when tested at die-level, with the microwave reflection spectrum shown in Fig.\,\ref{fig:MMQ} (trace A). The microwave Q measured here is most likely already limited by the dielectric and piezoelectric coupling induced loss by the residual TFLN traces supporting optical waveguides. In addition to the microwave loss induced by LN itself, the microwave Q is significantly impacted by the device packaging. For example, the wire bonds used for dc-tuning contribute directly to the microwave loss. As indicated by trace B shown in Fig.\,\ref{fig:MMQ}, when measured with wire bonds in the same closed box at \SI{2.6}{\kelvin}, the device displays a reduced intrinsic Q of $5.4 \times 10^3$.

In the final EO conversion package, additional microwave loss also arises from the cryogenic optical interfaces required to accommodate the fiber array. The fiber array is manipulated by a set of attocubes for precision alignment, preventing the use of a fully closed copper box. A copper box with open lid is utilized instead, similar to the configuration previously used in AlN microwave-to-optics experiment~\cite{han2020cavity} where we did not observe significant Q change before and after packaging. However, this is not the case for the TFLN converter studied here. As shown in Fig.\,\ref{fig:MMQ} (trace C), the device exhibits a further degradation of microwave quality factor, and the extracted intrinsic Q drops to $1.3 \times 10^3$. We conjecture that the excess microwave loss is mainly induced by the stronger piezoelectric coupling of LN (than AlN) to spurious modes supported by the RF packaging. To mitigate this loss, it is critical to further reduce the TFLN trace volume that does not participate in the EO conversion, and utilize a RF packaging that can effectively suppress spurious modes.


\section{bidirectional microwave-to-optics conversion}

\begin{figure}[!t]
\begin{centering}
\includegraphics[width=0.48\textwidth]{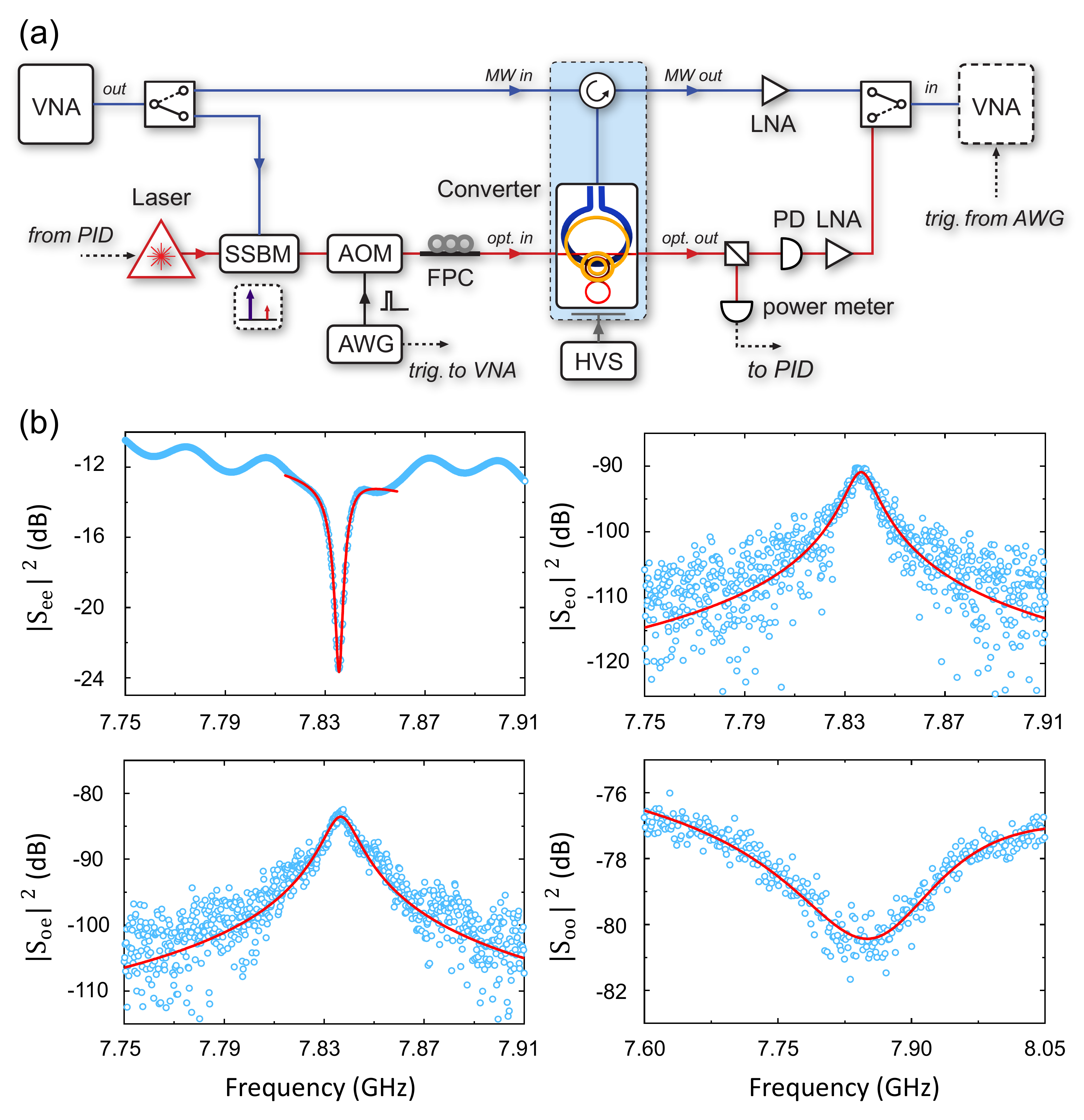}
\par\end{centering}
\caption{\label{fig:Smatrix} (a) Schematic of measurement setup. For CW measurement, the RF signal generated by a vector network analyzer (VNA) is sent to the device either as the microwave input or as the optical input through an optical single-sideband modulator (SSBM). The output from the device is then sent back to the VNA. For pulsed measurements, the pump is gated by an acoustic-optic modulator (AOM) driven by an arbitrary wave generator (AWG), which also triggers VNA. The pump is locked on resonance using a proportional–integral–derivative (PID) controller.  FPC: fiber polarization controller. PD: photodetector. LNA: low-noise amplifier. HVS: high voltage source.
(b) Measured microwave reflection $S_{\mathrm{ee}}$, optical-to-microwave conversion $S_{\mathrm{eo}}$, microwave-to-optical conversion $S_{\mathrm{oe}}$ and optical reflection $S_{\mathrm{oo}}$. The peak pump power is \SI{-12.0}{\dBm} in the waveguide and all matrix coefficients are normalized to the output power of VNA.}
\end{figure}

The schematic setup for EO conversion measurement is presented in Fig.\,\ref{fig:Smatrix}(a). The device is loaded in a closed-cycle refrigerator (Cryomech) cooled down to \SI{1.9}{\kelvin} base temperature. A tunable telecom laser (Santec 710) amplified by erbium-doped fiber amplifier (EDFA) is tuned to the anti-symmetric mode (lower frequency supermode) frequency as a parametric pump. In order to compensate for the residual PR-induced slow resonance frequency drift, 1\% of the transmitted light from the device is sampled to provide wavelength feedback to the laser through its build-in piezo. RF signal from a vector network analyzer (VNA) is either directly sent to the device as the microwave signal input or used to generate the optical signal input through optical single-sideband modulation~\cite{fan2018superconducting}. The microwave output from the converter is analyzed after a low-noise amplifier at room temperature. The optical output is collected and down-converted through heterodyne with the pump laser by a high-speed photodetector and then sent to the VNA. For measurement in high power regime, we reduce the thermal load to the device by gating the pump light to \SI{1}{\milli\second} pulses, using a high extinction acoustic-optic modulator (AOM) driven by an arbitrary waveform generator at 10\% duty cycle with a repetition of \SI{10}{\milli\second}. In the pulsed mode, each point of the VNA sweep is triggered by the AWG to synchronize with the pump pulse.

\begin{figure} [t]
\includegraphics[width=0.48\textwidth]{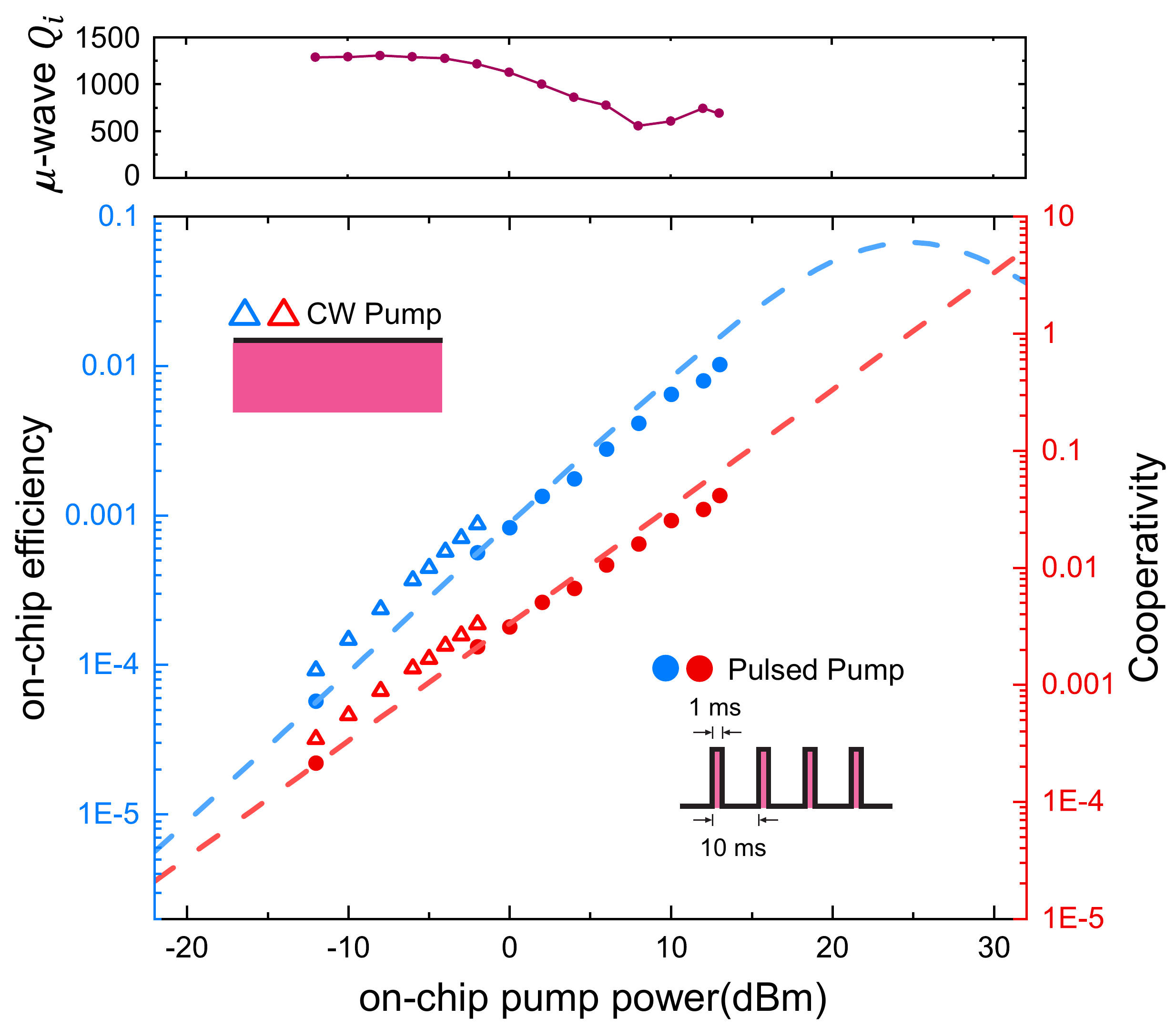}
\caption{\label{fig:efficiency} The on-chip conversion efficiency (blue) and cooperativity (red) as a function of on-chip peak pump power. The discontinuity in the CW and pulsed pump data is due to the use of different pairs of optical modes during the measurements. The maximum efficiency is recorded with a peak pump power of \SI{13.0}{\dBm}, reaching an on-chip efficiency of $1.02 \%$, internal efficiency of $15.2\%$ and cooperativity of $0.041$. The dashed lines are the linear predictions obtained from the data measured with a pulsed pump. The measured intrinsic microwave Q under different optical pump power is presented in the upper panel.}
\end{figure}

With a DC bias of \SI{220}{\volt}, the optical mode splitting is tuned (at a rate of $\sim$ \SI{0.6}{\pico\meter/\volt}) to match with the microwave resonance frequency. Fig.\,\ref{fig:Smatrix} shows the full spectra of scattering matrix elements employed for calibrating the conversion efficiency: optical refection $S_{\mathrm{oo}}$, microwave refection $S_{\mathrm{ee}}$, microwave-to-optical conversion $S_{\mathrm{oe}}$ and optical-to-microwave conversion $S_{\mathrm{eo}}$. The bidirectional nature of the EO conversion is thus fully established. The Lorentzian fit of the $S_{\mathrm{oe}}$ and $S_{\mathrm{eo}}$ spectral yield a \SI{3}{\decibel} bandwidth matching the microwave linewidth. The peak conversion efficiency is obtained by calibrating out the off-chip gain and loss factor of each input and output signal path\cite{andrews2014bidirectional}:
\begin{equation}
\label{eq4}
  \eta=\frac{S_{\mathrm{eo,p}}  S_{\mathrm{oe,p}} }{S_{\mathrm{oo,bg}} S_{\mathrm{ee,bg}} },
\end{equation}
\noindent where $S_{\mathrm{eo,p}}$ and $S_{\mathrm{eo,p}}$ are the peaks of conversion spectra, $S_{\mathrm{oo,bg}} $ and $S_{\mathrm{oo,bg}}$ are the backgrounds of reflection spectra, respectively.

We gradually increase the on-chip pump power from \SI{-12.0}{\dBm} to boost the on-chip conversion efficiency  which is obtained from the above-mentioned calibration method. Considering the loss and coupling characteristics of microwave and optical cavities, we could also extract the internal conversion efficiency $\eta_{\mathrm{int}}$ and thus cooperativity $C$ from Eqn.\,\ref{eq3} under different pump power, as shown in Fig.\,\ref{fig:efficiency}. We switch to 10\% duty cycle pulsed mode from \SI{-2.0}{\dBm} on-chip power to reduce the thermal load on the fridge. After extended exposure during CW measurement, the pair of optical modes we used are no longer accessible due to the accumulated charge. Therefore, we switch to another pair of optical modes in the pulsed measurement, leading to a discontinuity in the data trace. The detailed device characteristic is presented in the supplementary material.  
 
The maximum on-chip conversion efficiency of $1.02\%$ (internal efficiency of $15.2\%$) is extracted with a peak pump power adjusted to \SI{13.0}{\dBm} in the waveguides. The corresponding cooperativity reach a value of $0.041$, which is significantly higher than the recent reported values~\cite{mckenna2020cryogenic}. We note that the cooperativity no longer increases linearly with peak pump power in high power regime. This is attributed to the quality factor degradation of the superconducting resonator, as shown in the upper panel of Fig.~\ref{fig:efficiency}. Similar microwave loss has also been observed in previous studies~\cite{mckenna2020cryogenic, holzgrafe2020cavity}. The source of the additional microwave loss could be a combination of light absorption of the superconductor, thermo-optic heating, and undesired coupling from cryogenic packaging. From a linear fitting (dashed lines in Fig.\,\ref{fig:efficiency}(a) of the conversion efficiency in the low power regime, the vacuum coupling rate can be estimated to be $g_{eo}=2\pi \times 750 \mathrm{Hz}$, which is within the same order of magnitude compared to simulated value ($2\pi \times 1.5 \mathrm{kHz}$, see supplementary material). 



\section{Discussion, prospect and conclusion}

According to Eqn.~\ref{eq3}, the conversion performance of our current devices is mostly limited by the microwave Q affected by device packaging for cryogenic operation. The microwave Q, when measured on die-level, shows significantly higher value ($\sim10^4$) compared to prior works \cite{holzgrafe2020cavity,mckenna2020cryogenic}. However, in the final packaged conversion device, the measured intrinsic quality factor remains low compared to similar resonators patterned on other type of substrates, such as AlN~\cite{fan2018superconducting, fu2020ground}. With improved cryogenic packaging to suppress parasitic losses, if the intrinsic Q of $\sim10^4$ can be recovered, unitary internal conversion efficiency ($C=1$) can be obtained at the current pump power level. Also, the on-chip efficiency could be further improved by engineering the external loss rate of optical and microwave modes. Our current devices use grating couplers which introduce a insertion loss as high as \SI{12.0}{\decibel} per facet due to weak index contrast between LN and sapphire substrate. By ultilizing side-coupled inverted tapers, the insertion loss could be reduced to \SI{3.4}{\decibel} per facet \cite{fu2020ground,he2019low}, which will also dramatically reduce the thermal-optic heating and light absorption induced by scattered light. Finally, we note that the microwave resonator is not at its ground state (thermal photon occupancy $\bar{n}_\mathrm{th} \sim 5$) in this work. Future characterization of ground state conversion will be done in a dilution refrigerator at milikelvin temperature \cite{fu2020ground}. Using blue-detuned pump pulses and photon-counting setup, non-classical correlation and entanglement between microwave and optical photon can be demonstrated \cite{riedinger2018remote,riedinger2016non}.


In conclusion, we demonstrate that it is possible to mitigate the most detrimental PR effect and charge-screening effects in TFLN and achieve conversion efficiency up to $1.02\%$ maintaining stable device operation at cryogenic temperatures. 
Such stable operation at cryogenic temperature not only allows us to reach high conversion efficiency but also is fundamentally critical for quantum applications such as generation of entangled microwave and optical photon pairs, conversion of non-classical photon states from the microwave to optical domain along with entanglement transfer \cite{rueda2019electro}. Our device performance can be further improved by proper cryogenic packaging and light coupling. With such improvements, we anticipate to maintain high conversion efficiency with sufficiently low pump power for ground state operation at millikelvin temperature.  
 


~\\
\noindent\textbf{Funding.} This work is funded by ARO under grant number W911NF-18-1-0020. HXT acknowledges partial supports from NSF (EFMA-1640959), ARO (W911NF-19-2-0115) and the Packard Foundation. Funding for substrate materials used in this research was provided by DOE/BES grant under award number DE-SC0019406. 

~\\
\noindent\textbf{Acknowledgements.} The authors thanks Michael Rooks, Yong Sun, Sean Rinehart and Kelly Woods for support in the cleanroom and assistance in device fabrication. 

~\\
\noindent\textbf{Disclosures.} The authors declare no conflicts of interest.

\bibliography{conv_ref}

\end{document}